\documentclass[journal=jacsat,manuscript=article]{achemso}
\usepackage{graphicx}%[draft]
\usepackage[intlimits,tbtags]{amsmath}
\usepackage[dvipsnames]{xcolor}
\usepackage{amssymb}
\usepackage{array}
\usepackage{dcolumn}
\usepackage{bm}
\usepackage{multirow}
\usepackage{siunitx}
\usepackage[version=4]{mhchem}

\newcolumntype{d}[1]{D{.}{.}{#1} }

\graphicspath{ {./Figs/} {./} }

\newcommand{\jena}{Institut f\"ur Festk\"orpertheorie und -optik,
                   Friedrich-Schiller-Universit\"at Jena and European Theoretical Spectroscopy Facility,
                   Max-Wien-Platz 1, 07743 Jena, Germany}

\newcommand{\halle}{Institut f\"ur Physik, Martin-Luther-Universit\"at
  Halle-Wittenberg, D-06099 Halle, Germany}
  
\newcommand{\vienna}{Institute of Materials Chemistry, Vienna University of Technology,
                     Getreidemarkt 9/165-TC, A-1060 Vienna, Austria}

\author{Pedro Borlido}
\affiliation{\jena}
\author{Jan Doumont}
\affiliation{\vienna}
\author{Fabien Tran}
\affiliation{\vienna}
\author{Miguel A. L. Marques}
\email{miguel.marques@physik.uni-halle.de}
\affiliation{\halle}
\author{Silvana Botti}
\email{silvana.botti@uni-jena.de}
\affiliation{\jena}

\title{Validation of pseudopotential calculations for the electronic band gap of solids}

\begin{document}

%\title{Validation of pseudopotential calculations for the electronic band gap of solids}
%\author{Pedro M. Borlido}
%\affiliation{\jena}
%\author{Jan Doumont}
%\affiliation{\vienna}
%\author{Fabien Tran}
%\affiliation{\vienna}
%\author{Silvana Botti}
%\email{silvana.botti@uni-jena.de}
%\affiliation{\jena}
%\author{Miguel A.~L. Marques}
%\email{miguel.marques@physik.uni-halle.de}
%\affiliation{\halle}

\begin{abstract}
%%%%%%%%%%%%%
%%% Do not use abbreviations in the abstract - SB
Nowadays pseudopotential density-functional theory calculations constitute the standard approach to tackle solid-state electronic problems.
These rely on distributed pseudopotential tables that were built from all-electron atomic calculations using few popular semi-local exchange-correlation functionals, while pseudopotentials based on more modern functionals, like meta-GGA and hybrid functionals, or for many-body methods, such as $GW$, are often not available.
Because of this, employing pseudopotentials created with inconsistent exchange-correlation functionals has become a common practice.
Our aim is to quantify systematically the error in the determination of the electronic band gap when cross-functional pseudopotential calculations are performed. To this end we compare band gaps obtained with norm-conserving pseudopotentials or the projector-augmented wave method with all-electron calculations for a large dataset of 473 solids. We focus in particular on density functionals that were designed specifically for band-gap calculations.
On average, the absolute error is about 0.1~eV, yielding absolute relative errors in the 5-10\% range.
Considering that typical errors stemming from the choice of the functional are usually larger, we conclude that the effect of choosing an inconsistent pseudopotential is rather harmless for most applications.
However, we find specific cases where absolute errors can be larger than 1~eV, or others where relative errors can amount to a large fraction of the band gap.

\end{abstract}
% PACS
%\pacs{}

\maketitle

\section{Introduction}
\label{sec:introduction}

% > KS-DFT
% > XC-correlation
Since its origin more than 50 years ago, density functional theory~\cite{hohenbergPR1964,kohnPR1965} (DFT) has become the standard approach to tackle the electronic structure of solids.
A workable approach to DFT is attained via the Kohn-Sham formulation~\cite{kohnPR1965}, leading to equations that can be solved
efficiently with modern computational resources. Although DFT is in principle exact, it relies on approximations of the exchange-correlation (xc) energy functional that inherently limit the accuracy of calculations~\cite{KuemmelRMP08,cohenCR2012,BurkeJCP12,BeckeJCP14}.

% > All electronccccccccccccccccccc
From a purely theoretical point of view, the xc functional is the only approximation in DFT. However, in practice, the Kohn-Sham equations -- a system of coupled, 
non-linear, partial differential equations -- are solved numerically, introducing further approximations. Several different approaches have been developed and are commonly used in the physics and chemistry communities. A basic distinction exists between all-electron and pseudopotential (or effective-core potential) methods. In the former, all electrons are explicitly included in the calculation, and the electron-nuclear attraction is described by the standard Coulomb potential. The latter method is based on the distinction between core and valence electrons. By replacing the 
effect of the nucleus and the core electrons by an effective pseudopotential, efficient plane-wave~\cite{pickettCPR1989} or real-space~\cite{andradePCCP2015, enkovaaraJPCM2010}
techniques can be used to solve the Kohn-Sham equations. As a middle way, the projector-augmented wave method (PAW)~\cite{blochlPRB1994,kressePRB1999} was developed, combining the advantages of pseudopotentials with a reconstruction of the all-electron wavefunction. 

% > Pseudo potentials
In materials science DFT calculations are often performed with effective-core methods~\cite{Martin,LejaeghereaadS16}. It is generally believed that well-tuned pseudopotentials can yield very precise calculations of many properties of solids at a much lower cost than all-electron methods. However, in order to guarantee this, careful optimization is necessary to ensure the quality and transferability of the pseudopotential for the situation at hand. 
Several well-tested tables are available to the community, but all these pseudopotentials are built based on two approximations to the xc energy functional:
the local density approximation (LDA)~\cite{VoskoCJP80,perdewPRB1981,perdewPRB1992,perdewPRB2018})
and the generalized-gradient approximation (GGA), and more specifically the Perdew-Burke-Ernzerhof (PBE)~\cite{perdewPRL1996} functional, which is the \textit{de facto} standard in the physics community.
This is true for norm-conserving~\cite{hamannPRL1979} and ultra-soft~\cite{vanderbiltPRB1990} potentials, as well as for the PAW method~\cite{blochlPRB1994}. 
The disparity between the number of functionals in generally available pseudopotential datasets (LDA, PBE and sometimes PBEsol~\cite{perdewPRL2008})
and the number of available xc functionals in the literature~\cite{lehtolaS2018,marquesCPC2012} (more than 500) is therefore confounding, especially if we consider that \textit{ab initio} pseudopotentials have been well established since the late 70's~\cite{hamannPRL1979, zungerPRB1978} and that recently developed xc functionals are acknowledged to be more accurate for electronic structure calculations than standard LDA and GGA functionals~\cite{sunPRL2015, adamoTJoCP1999,perdewTJoCP1996, heydTJoCP2003,krukauTJoCP2006}.
As a result of this situation, the great 
majority of calculations performed with improved functionals makes use of an inconsistent pseudopotential, built from LDA or GGA atomic calculations. This problem is relevant for both total-energy and band-structure calculations. For the latter, which are the focus of the present work, the most accurate calculations use either hybrid functionals (such as the Heyd-Scuseria-Ernzernhof 2006 functional~\cite{heydTJoCP2003,krukauTJoCP2006}), $GW$ methods~\cite{HedinPR65,HybertsenPRB86} or specialized semilocal functionals, such as the modified Becke-Johnson potential~\cite{tranPRL2009} (mBJLDA).
For example, despite the publication of several works on the generation of pseudopotentials for Hartree-Fock and hybrid functionals~\cite{al-saidiPRB2008,trailTJoCP2005,xuPRB2010,yangPRB2018,TanPCCP18,TanJCP19}, the lack of 
readily available tables means that most of these calculations are done with PBE ones.
Furthermore, some codes do not give the user the possibility of changing the pseudopotentials. % PB: do we really need this sentence?
The situation is even more complicated for the mBJLDA potential, as it is defined for periodic systems and can not be applied to atoms, unless specific schemes are used~\cite{BartokPRB19,Rauch2020}. Note that a similar problem exists also for many-body $GW$~\cite{HedinPR65,HybertsenPRB86} or LDA+$U$~\cite{anisimov1997first,anisimov1991density} calculations, that are often performed with LDA or PBE pseudopotentials.

This common practice creates an inconsistency, which invariably introduces some uncontrollable errors in the calculations. Even if this is well-known, relatively little attention has been given to quantify the effect of cross-functional pseudopotential calculations~\cite{yaoTJoCP2017, fuchsPRB1998, yangPRB2018,BartokPRB19}. With this in mind, we decided to study the error coming from using an inconsistent pseudopotential in the calculation of band gaps. This was done by comparing the results obtained with three codes: {\sc WIEN2K}~\cite{WIEN2k,BlahaJCP20}, an all-electron code that uses the augmented 
plane wave plus local orbitals (APW+lo) method~\cite{singh2006}, {\sc ABINIT}~\cite{gonzeCPC2016,GONZE2020107042} with norm-conserving pseudopotentials, and {\sc VASP}~\cite{kressePRB1996,kressePRB1999} with PAW setups.
Note that our comparison goes beyond the $\Delta$-test~\cite{LejaeghereaadS16}, as we compare band gaps, that also include the influence of unoccupied bands, and we take into account errors coming from using inconsistent pseudopotentials.

\section{Methods}
\label{sec:methods}

\subsection{Dataset}
% > Structures
We performed calculations for all materials contained in the dataset developed in Ref.~\cite{borlidoJCTC2019}, that counts 473 non-magnetic semiconductors. This dataset covers the majority of the periodic table, and includes materials with a wide range of band gaps. All calculations were performed at the experimental geometry (see Ref.~\cite{borlidoJCTC2019} for more details on the materials dataset).

\subsection{Functionals}
% > Functionals

Besides the standard LDA and PBE functionals, our choice of xc functionals (or potentials) was based on two criteria: (i)~their availability in all three software packages we used (see Sec.~\ref{sec:codes} for more information on the codes) and 
(ii)~their relevance for the calculation of band gaps.
Our final choice includes the Perdew-Wang (PW92)~\cite{perdewPRB1992} and the Perdew-Zunger (PZ81)~\cite{perdewPRB1981} parametrization of the LDA correlation, the local Slater potential (SLOC)~\cite{finzelIJQC2017}), which is a simple modification of the LDA exchange to approximate the Slater potential. We consider also several GGA functionals: PBE~\cite{perdewPRL1996}, revised PBE (RPBE)~\cite{hammerPRB1999}, Engel-Vosko (EV93)~\cite{engelPRB1993} combined with Perdew-Wang (PW91) correlation~\cite{perdewPRB1992a} and high-local exchange (HLE16)~\cite{vermaJPCL2017}.
Finally we used the meta-GGA mBJLDA~\cite{tranPRL2009}.
The PW92, PZ81 and PBE functionals are general-purpose approximations with
widespread use in the solid-state community for total-energy calculations. The mBJLDA potential is known to yield excellent band gaps, which is also the case of the simpler SLOC and HLE16, although they are slightly inferior to mBJLDA~\cite{tranJPCA2017,borlidoJCTC2019}. Note that meta-GGA energy functionals are not supported self-consistently in the reference APW+lo {\sc WIEN2k} code, and therefore we did not include in our analysis the recent meta-GGAs HLE17~\cite{VermaJPCC17} and TASK~\cite{AschebrockPRR19}, which also perform very well for band gaps. Hybrid functionals are available in the three considered codes, however they lead to calculations which are several orders of magnitude more expensive, in particular if parameters for highly converged calculations are used. Therefore, since our test set is very large, we refrained from using hybrid functionals in the present work. For the same reason we did not consider $GW$ methods.

\subsection{Codes}
\label{sec:codes}
% > WIEN2K calculations
The all-electron calculations were done with the {\sc WIEN2k} package~\cite{WIEN2k,BlahaJCP20}, which uses the APW+lo basis set~\cite{singh2006}. The calculations were done with sufficiently large parameters (e.g., basis-set size) to ensure convergence of band gaps within $\sim0.05$~eV. The {\sc WIEN2k} results will serve as reference for the comparison with the other codes.

% > VASP calculations
The PAW calculations were performed using a custom version of the Vienna ab initio simulation package ({\sc VASP}; version 5.4)~\cite{kressePRB1996,kressePRB1999} that is interfaced to {\sc LIBXC}~~\cite{lehtolaS2018,marquesCPC2012}. A plane wave cutoff of 520~eV was used for all species along with the same \textbf{k}-grids as in Ref.~\cite{borlidoJCTC2019}.
All meta-GGA calculations were performed accounting for non-spherical contributions of the density gradient inside the augmentation spheres.
{\sc VASP} calculations are in general restricted to the PAW sets included in the distribution, and we were therefore able to use PBE and LDA PAW pseudopotentials.

% > ABINIT calculations
The {\sc ABINIT} package~\cite{gonzeCPC2016} was used to test norm-conserving pseudopotential calculations.
For LDA and PBE pseudopotentials we resorted to the Pseudo Dojo distribution~\cite{vansettenCPC2018} (version 0.4, with stringent accuracy). Although this set
covers most of the periodic table, some elements like thorium are absent and had unfortunately to be left out of the calculations. 
Since {\sc ABINIT } does not currently support pseudopotentials with non-linear core corrections for meta-GGAs, mBJLDA calculations with this code were not performed. 
For SLOC and HLE16 we generated a set of pseudopotentials using the {\sc ONCVPSP} package~\cite{hamannPRB2013} (version 3.3.1).
As a starting point we used the input files from the stringent set of the Pseudo Dojo distribution, changing them to include the desired functionals.
Whenever a particular input leads to unsuccessful/difficult calculations, we exchanged it for that of the standard set.
Further small changes to the local part of the pseudopotential were performed in order to avoid spurious effects (e.g. ghost states) that were detected by the post-processing tools of {\sc ONCVPSP}.
We note that although care was taken in this process, we did not perform further tests ($\Delta$-test, GBVR test, etc.~\cite{garrityCMS2014,jolletCPC2014,lejaeghereCRiSSaMS2014}) or optimizations.
Therefore, and although these pseudopotentials yield generally consistent results for band gaps, care should be taken for a more general use.
The pseudopotentials are given as Supplementary Information, and the whole set can also be downloaded from~\cite{tddft}.

Note that the three codes, as well as {\sc ONCVPSP}, are linked to the library of xc functionals {\sc LIBXC}~\cite{lehtolaS2018,marquesCPC2012}, which allows to access several hundreds of functionals, including the ones considered here.

All calculations were performed neglecting spin-orbit coupling.
This term is expected to contribute on average about 0.1~eV to the band gap. This amount is considerably smaller than the typical average error of the xc functionals. Anyway, as all calculations were performed consistently without this term, its exclusion does not affect our comparison between the codes, that is the main purpose of the present work.

\subsection{Statistics}
For the analysis of the results we compare band gaps calculated with norm-conserving pseudopotentials and PAW methods to all-electron values.
Our analysis is restricted to those materials that were not determined by {\sc WIEN2k} to have a theoretical band gap smaller than 0.01~eV, despite being measured to be semiconductors. From the original 473 entries, {\sc WIEN2k} predicts between 13 and 40 of such materials, depending on the functional.

Whenever presenting the results we use the notation \texttt{<calculation xc>@<pseudo xc>}, and if the code used is ambiguous, we precede this string with its name.
For example, {\sc ABINIT}:PBE@LDA would make reference to the set of values computed with {\sc ABINIT}, using the PBE calculations with LDA pseudopotentials.

The statistical analysis is based on the determination of
the mean absolute error, $\text{MAE} = \sum^n_{i=1} |y_i - y_{i, \text{exp}} |/n$;
the mean error, $\text{ME} = \sum^n_{i=1} (y_i - y_{i, \text{exp}} )/n $;
the standard deviation of the errors, $\sigma = \sqrt{ \sum^n_{i=1} (y_i - y_{i, \text{exp}} - \text{ME} )^2/n }$;
the median error (MnE);
the interquartile range (IQR);
the median of the absolute deviations from the median (MADM);
the mean absolute percentage error, $\text{MAPE} = 100 \times \sum^n_{i=1} |y_i - y_{i, \text{exp}} |/( n \, y_{i, \text{exp}} )$; the mean percentage error $\text{MPE} = 100 \times \sum^n_{i=1} (y_i - y_{i, \text{exp}} )/( n \, y_{i, \text{exp}} )$; and the maximum absolute error in the calculation of band gaps with respect to experimental values.
The complete set of results is presented in Tables~SI--SVII of the Supplementary Information, while Table~\ref{tab:standard-summary} shows a summary of the most important statistical quantities.

After a preliminary analysis of the results, it became apparent that relative quantities (such as the MAPE) were being extremely affected by materials with very small band gaps. This is easy to understand 
as small errors lead to a rather large relative error for systems with band gaps in the range 0.1--0.2~eV, skewing significantly the statistical averages. Therefore we opted to 
consider in Table~\ref{tab:standard-summary} percentage quantities (MAPE and MPE) for the subset of systems with band gaps larger than $0.25$~eV.
Absolute quantities were still computed for the entire dataset.

\section{Results}
\label{sec:results}

\begin{figure*}
    \centering
    \includegraphics[width=0.35\textwidth]{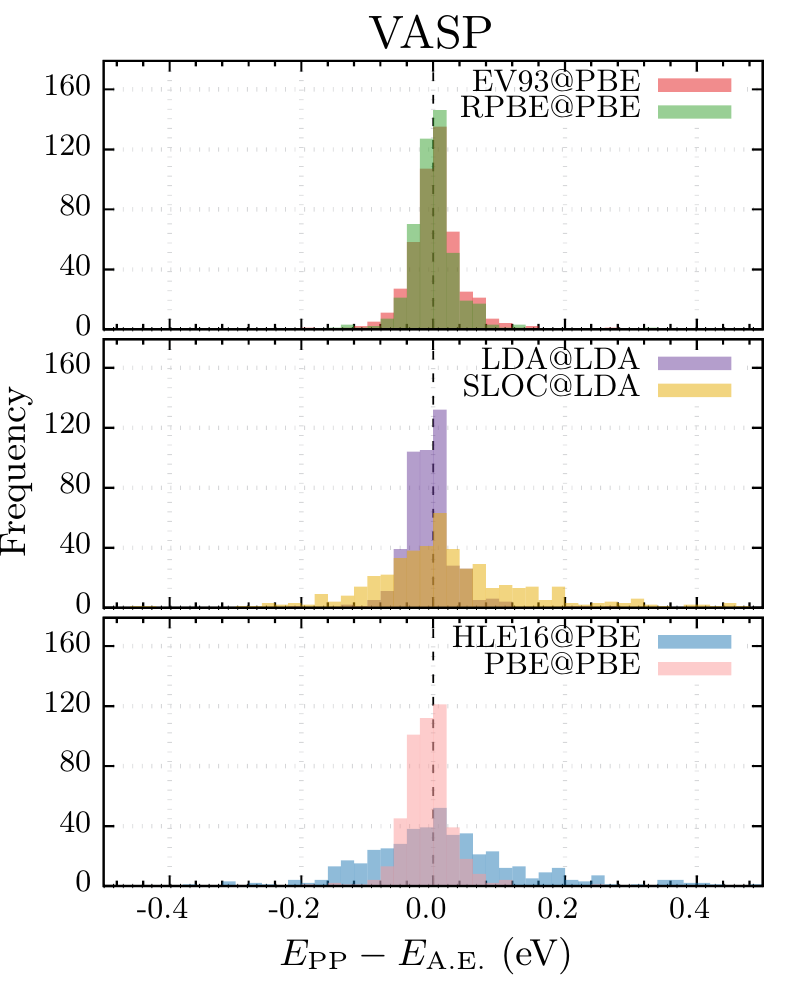}
    \includegraphics[width=0.35\textwidth]{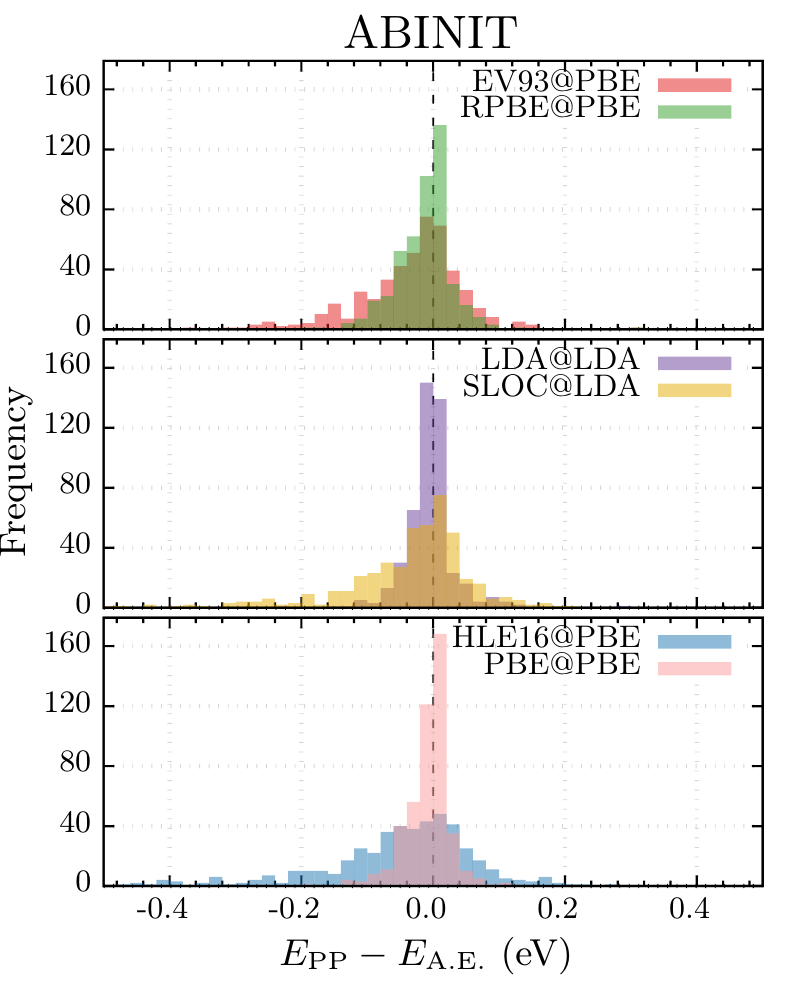}
    \caption{Histograms of errors (with respect to the all-electron results) for band gaps computed with {\sc VASP} (left) and {\sc ABINIT} (right). Boxes have a width of $0.02$~eV.}
    \label{fig:errors-std}
\end{figure*}

We start our analysis by looking at the results computed with the generally available LDA and PBE datasets.
As visible in Table~\ref{tab:standard-summary}, band-gap calculations performed with LDA and PBE xc functionals on top of the corresponding pseudopotentials are in excellent agreement with all-electron calculations. Not only are the MAE and ME for these calculations very small (in absolute value smaller than $0.03$~eV), but also the dispersion
of results is quite localized. This is visually represented in the error histograms of Fig.~\ref{fig:errors-std}, and also in the corresponding standard
deviation $\sigma$ (smaller than $0.06$~eV).
The maximum absolute error in these conditions is around 0.2--0.3~eV. However, one can see from the distribution of errors that absolute values larger
than 0.1~eV are rare.
These conclusions are valid for both pseudopotential and PAW calculations, although the MPE and MAPE of the latter are consistently larger. This may be explained by the fact that the Pseudo Dojo sets~\cite{vansettenCPC2018} are much more recent than the PAW sets available in {\sc VASP}, and that they were systematically optimized. In any case, these errors are certainly acceptable for the large majority of applications. This confirms the generally accepted idea that effective-core methods (either norm-conserving or PAW approach) are reliable for the calculations of band gaps.

{
\renewcommand{\arraystretch}{1.2}
\newcolumntype{L}[1]{>{\raggedright\let\newline\\\arraybackslash\hspace{0pt}}m{#1}}
\newcolumntype{C}[1]{>{\centering\let\newline\\\arraybackslash\hspace{0pt}}m{#1}}
\newcolumntype{R}[1]{>{\raggedleft\let\newline\\\arraybackslash\hspace{0pt}}m{#1}}
\begin{table*}
   \centering
   \caption{M(A)E (in eV), standard deviation (in eV), M(A)PE (in \%, for band gaps larger than 0.25~eV), and maximum absolute error (in eV) with respect to the all-electron results obtained with the considered functionals when a standard (LDA or PBE) pseudopotential is used.}
   \label{tab:standard-summary}
   \begin{tabular}{R{4cm} C{1.2cm} C{1.2cm} C{1.2cm} C{1.2cm} C{1.5cm}  C{2.8cm}  } \hline \hline
                       &    ME  & MAE & $\sigma$ & MPE & MAPE  &  Max. Err.            \\ \hline
      \multicolumn{3}{l}{{\sc ABINIT} }                                   &                \\
      LDA@LDA          &   $ 0.01$   & 0.02 &   0.04    &  0.45  &    2.47    &  -0.28 (\ce{SnTe})        \\
      SLOC@LDA         &   $ 0.05$   & 0.08 &   0.14    &  3.13  &    5.43    &   1.33 (\ce{LaF3})        \\
      PBE@PBE          &   $ 0.01$   & 0.02 &   0.06    &  0.82  &    2.31    &   0.19 (\ce{GeAs})        \\
      RPBE@PBE         &   $ 0.01$   & 0.03 &   0.04    &  0.98  &    2.89    &  -0.30 (\ce{SnTe})        \\
      EV93@PBE         &   $ 0.03$   & 0.06 &   0.08    &  2.26  &    4.51    &   0.36 (\ce{ZnO})         \\
      HLE16@PBE        &   $ 0.06$   & 0.10 &   0.14    &  3.57  &    6.50    &   1.10 (\ce{LaF3})        \\
      \multicolumn{3}{l}{{\sc VASP} }                                     &                \\
      LDA@LDA          &   $ 0.01$   & 0.03 &  0.04     &  0.93     &    2.96    &  -0.27 (\ce{SnTe})        \\
      SLOC@LDA         &   $-0.03$   & 0.09 &  0.14     &  $-1.48$  &    5.76    &  -0.76 (\ce{BiF3})        \\
      mBJLDA@LDA       &   $-0.08$   & 0.11 &  0.26     &  $-2.06$  &    4.54    &  -3.87 (\ce{Ne})           \\
      PBE@PBE          &   $ 0.01$   & 0.03 &  0.04     &  1.17     &    2.74    &  -0.25 (\ce{LaF3})         \\
      RPBE@PBE         &   $ 0.00$   & 0.02 &  0.04     &  0.16     &    2.33    &  -0.32 (\ce{SnTe})         \\
      EV93@PBE         &   $ 0.00$   & 0.03 &  0.04     &  $-0.18$  &    2.24    &  -0.28 (\ce{SnTe})         \\
      HLE16@PBE        &   $-0.03$   & 0.11 &  0.17     &  $-1.75$  &    6.12    &  -1.17 (\ce{NaCl})         \\
      \hline \hline
   \end{tabular}
\end{table*}
}

{
\renewcommand{\arraystretch}{1.1}
\newcolumntype{L}[1]{>{\raggedright\let\newline\\\arraybackslash\hspace{0pt}}m{#1}}
\newcolumntype{C}[1]{>{\centering\let\newline\\\arraybackslash\hspace{0pt}}m{#1}}
\newcolumntype{R}[1]{>{\raggedleft\let\newline\\\arraybackslash\hspace{0pt}}m{#1}}
\begin{table*}
   \centering
   \caption{List of the ten materials with highest absolute error (with respect to the all-electron values) in the band gap (in eV) obtained with the  {\sc ABINIT} code for the SLOC and HLE16 functionals. In parentheses is indicated the all-electron band gap of the corresponding material (in eV). }
   \label{tab:list-mats-error-abinit}
   \begin{tabular}{ C{2.4cm}  C{2.2cm} | C{2.4cm}  C{2.2cm}   } \hline \hline
\multicolumn{4}{c}{  {\sc ABINIT} } \\ \hline
\multicolumn{2}{c}{SLOC@LDA}  & \multicolumn{2}{c}{HLE16@PBE}   \\
Material   & Error    &  Material   & Error     \\ \hline
\ce{AlPO4}      &    0.47   (7.75) &  \ce{LaCuOTe}    &    0.44   (0.65)   \\
\ce{SiO2}       &    0.48   (7.70) &  \ce{ZnSnO3}     &    0.44   (2.47)   \\
\ce{LaCuOSe}    &    0.51   (0.44) &  \ce{GeO2}       &    0.45   (2.12)   \\
\ce{Al2O3}      &    0.52   (7.25) &  \ce{ZnO}        &    0.48   (2.82)   \\
\ce{LaCuOS}     &    0.55   (0.57) &  \ce{LaCuOSe}    &    0.52   (0.99)   \\
\ce{K2La2Ti3O10}&    0.60   (1.17) &  \ce{LaCuOS}     &    0.56   (1.17)   \\
\ce{La2O3}      &    0.85   (1.89) &  \ce{NaLa2TaO6}  &    0.71   (2.63)   \\
\ce{NaLa2TaO6}  &    0.85   (1.73) &  \ce{CuLaO2}     &    0.76   (2.38)   \\
\ce{CuLaO2}     &    0.91   (1.75) &  \ce{La2O3}      &    0.77   (2.83)   \\
\ce{LaF3}       &    1.33   (2.59) &  \ce{LaF3}       &    1.10   (3.81)   \\
\hline \hline
   \end{tabular}
\end{table*}
}

{
\renewcommand{\arraystretch}{1.1}
\newcolumntype{L}[1]{>{\raggedright\let\newline\\\arraybackslash\hspace{0pt}}m{#1}}
\newcolumntype{C}[1]{>{\centering\let\newline\\\arraybackslash\hspace{0pt}}m{#1}}
\newcolumntype{R}[1]{>{\raggedleft\let\newline\\\arraybackslash\hspace{0pt}}m{#1}}
\begin{table*}
   \centering
   \caption{List of the ten materials with highest absolute error (with respect to the all-electron values) in the band gap (in eV) obtained with the {\sc VASP} code for the SLOC, HLE16 and mBJLDA functionals. In parentheses is indicated the all-electron band gap of the corresponding material (in eV). }
   \label{tab:list-mats-error-vasp}
   \begin{tabular}{ C{2.2cm}  C{2.2cm} | C{2.2cm}  C{2.2cm} | C{2.2cm}  C{2.6cm} } \hline \hline
\multicolumn{6}{c}{  {\sc VASP} } \\ \hline
\multicolumn{2}{c}{SLOC@LDA}  & \multicolumn{2}{c}{HLE16@PBE}    &    \multicolumn{2}{c}{mBJLDA@LDA}    \\
Material   & Error    &  Material   & Error   &  Material   & Error  \\ \hline
\ce{KTaO3}      &    -0.43  (2.35) &  \ce{ThO2}       &    -0.56  (3.83)  &  \ce{RbF}     &       -0.71  (9.49)\\
\ce{LiTaO3}     &    -0.44  (3.16) &  \ce{LiCoO2}     &    -0.59  (1.18)  &  \ce{LiF}     &       -0.73  (12.81)\\
\ce{BaCl2}      &    0.44   (5.86) &  \ce{BiF3}       &    -0.60  (4.49)  &  \ce{KCl}     &       -0.79  (8.56)\\
\ce{Ba2InTaO6}  &    -0.45  (4.08) &  \ce{La2O3}      &    -0.63  (2.83)  &  \ce{SrF2}    &       -0.86  (11.07)\\
\ce{NaTaO3}     &    -0.45  (2.75) &  \ce{NaLa2TaO6}  &    -0.63  (2.63)  &  \ce{AlPO4}   &       -0.87  (9.14)\\
\ce{HfO2}       &    -0.47  (4.31) &  \ce{NaBr}       &    -0.91  (6.45)  &  \ce{MgF2}    &       -0.90  (11.42)\\
\ce{AgF}        &    0.52   (3.07) &  \ce{NaF}        &    -0.93  (9.35)  &  \ce{NaF}     &       -1.34  (11.59)\\
\ce{Bi2O2CO3}   &    -0.57  (0.97) &  \ce{LaF3}       &    -1.01  (3.81)  &  \ce{Ar}      &       -1.35  (13.90) \\
\ce{LaF3}       &    -0.67  (2.59) &  \ce{NaI}        &    -1.10  (5.69)  &  \ce{LiIO3}   &       -1.75  (4.92) \\
\ce{BiF3}       &    -0.76  (4.70) &  \ce{NaCl}       &    -1.17  (7.63)  &  \ce{Ne}      &       -3.87  (22.72) \\
\hline \hline
   \end{tabular}
\end{table*}
}

Moving to the cross-functional calculations, we have to distinguish two different cases.
Rather good results are still found for RPBE@PBE and EV93@PBE, which is probably due to the fact that the functional form of RPBE and EV93 does not differ too much from PBE, so that the PBE pseudopotentials are still accurate for RPBE and EV93.
The increase in the mean errors is the largest for EV93@PBE with {\sc ABINIT}, 0.06~eV for the MAE and 4.51\% for the MAPE.

The situation is different for functionals that give accurate band gaps (HLE16, SLOC, and mBJLDA). Indeed, we can see a clear drop in accuracy, as seen by the increment in the various mean errors. For instance, the MAPE lies between 4.5\% and 6.5\% with both {\sc ABINIT} and {\sc VASP}, which is two or three times larger than when LDA/PBE calculations are done with a consistent pseudopotential. Note that these are still relatively small values, clearly smaller than the average errors with respect to experiment, which are of the order of 30\%~\cite{borlidoJCTC2019}. The loss in performance is more
strikingly seen in terms of the dispersion of the errors (see Fig.~\ref{fig:errors-std}). In particular, the standard deviation $\sigma$ reaches values in the range $0.14-0.26$~eV, which represents a threefold increase. This increase in dispersion leads inevitably to an increase in the maximum error, that can reach 1~eV or more as shown in Tables~\ref{tab:list-mats-error-abinit} and \ref{tab:list-mats-error-vasp}. For example, {\sc ABINIT}:SLOC@LDA and {\sc ABINIT}:HLE16@PBE yield errors of 1.33 and 1.10~eV for \ce{LaF3}, respectively, corresponding to very large relative differences of $\sim50$\% and $\sim30\%$ with respect to {\sc WIEN2k} values. {\sc VASP}-PAW setups do not perform better. Maximum errors of $-1.17$~eV with {\sc VASP}:HLE16@PBE (for \ce{NaCl}) and $-0.76$~eV with {\sc VASP}:SLOC@LDA (for \ce{BiF3}) are obtained. The situation is particularly worrying for mBJLDA@LDA, where we find a maximum deviation of $-3.87$~eV (for \ce{Ne}) with respect to {\sc WIEN2k}, while three other error values are above 1~eV.

{
\newcolumntype{L}[1]{>{\raggedright\let\newline\\\arraybackslash\hspace{0pt}}m{#1}}
\newcolumntype{C}[1]{>{\centering\let\newline\\\arraybackslash\hspace{0pt}}m{#1}}
\newcolumntype{R}[1]{>{\raggedleft\let\newline\\\arraybackslash\hspace{0pt}}m{#1}}

\begin{table}
   \centering
   \caption{MAE (in eV), standard deviation $\sigma$ (in eV) and MAPE (in \%) with respect to the all-electron results for the band gaps obtained using the {\sc ABINIT} code for all cross-functionals pseudopotential (PP) combinations.}
   \label{tab:pseudo-summary}
   \begin{tabular}{r r  C{1.2cm} C{1.2cm} C{1.4cm} C{1.2cm} } \hline  \hline
                      &         &  \multicolumn{4}{c}{Functional} \\
     & &  LDA & PBE & HLE16 & SLOC \\
    \hline
       & &        \multicolumn{4}{c}{MAE}                         \\
  \multicolumn{1}{ c }{ \multirow{14}{*}{PP} } & LDA & 0.02 & 0.05 & 0.12 & 0.08 \\
  \multicolumn{1}{ c }{} & PBE & 0.05 & 0.02 & 0.10 & 0.09 \\
  \multicolumn{1}{ c }{} & HLE16  & 0.10 & 0.07 & 0.03 & 0.08 \\
  \multicolumn{1}{ c }{} & SLOC & 0.08 & 0.07 & 0.10 & 0.03 \\
   & & \multicolumn{4}{c}{$\sigma$}   \\
                         & LDA & 0.04 & 0.06 & 0.16 & 0.14 \\
  \multicolumn{1}{ c }{} & PBE & 0.06 & 0.03 & 0.14 & 0.16 \\
  \multicolumn{1}{ c }{} & HLE16 & 0.14 & 0.11 & 0.04 & 0.12 \\
  \multicolumn{1}{ c }{} & SLOC & 0.13 & 0.13 & 0.13 & 0.05 \\
    & & \multicolumn{4}{c}{MAPE}   \\
                         & LDA & 2.47 & 4.88 & 7.79 & 5.43 \\
  \multicolumn{1}{ c }{} & PBE & 4.92 & 2.31 & 6.50 & 5.94 \\
  \multicolumn{1}{ c }{} & HLE16 & 8.92 & 5.87 & 2.22 & 5.37 \\
  \multicolumn{1}{ c }{} & SLOC & 6.90 & 6.50 & 6.28 & 1.94 \\
  \hline  \hline
   \end{tabular}
\end{table}
}

The solution to avoid such inaccuracies when using a new functional is rather clear and straightforward, namely a pseudopotential has to be generated, optimized and tested specifically for the new functional.
This is what we have done for HLE16 and SLOC. The results are shown in Table~\ref{tab:pseudo-summary}, where we list the MAE, standard deviation $\sigma$, and MAPE for all functional combinations of LDA, PBE, HLE16, and SLOC. As expected, using a LDA potential in a PBE calculation or vice-versa has relatively small effect on the overall quality of the band gaps. However, using an inconsistent potential in a SLOC or HLE16 calculation has a much larger effect (by a factor of at least three), confirming what was already discussed above. 
Since, grossly speaking, the difference between LDA and PBE results is smaller than the one between PBE/LDA and HLE16/SLOC, it is not entirely unexpected that an inconsistent pseudopotential for these functionals yields worse results.
Performing consistent calculations brings back the pseudopotential error to the normal range (below $0.05$~eV for the MAE and $\sigma$, and at $\sim 2\%$ for the MAPE). Unfortunately, generating consistent pseudopotentials is far from obvious for more sophisticated methods like hybrid functionals, many-body theories, or even the meta-GGA mBJLDA potential.

As the mBJLDA is currently the best performing semilocal functional for the prediction of band gaps~\cite{borlidoJCTC2019, tranJPCA2017}, a more detailed discussion is in order. The exchange component of mBJLDA (the correlation component consists of LDA) is essentially a rescaling of the Becke-Johnson~\cite{beckeJCP2006} exchange potential via a density-dependent parameter $c$:
\begin{equation}
 v^\text{mBJ}_{\text{x}}(\mathbf{r}) = c v^\text{BR}_{\text{x}}(\mathbf{r}) + (3c-2)\frac{1}{\pi}\sqrt{\frac{5}{6}}\sqrt{\frac{\tau(\mathbf{r})}{n(\mathbf{r})}} \, ,
	\label{eq:vx_mbj}
\end{equation}
where
\begin{equation}
c =
\alpha +
\beta
\left[
\frac{1}{V_{\text{cell}}}
\int_{\text{cell}}
\frac{ | \nabla n (\mathbf{r}') | }{ n (\mathbf{r}') }
\:
\mathrm{d} \mathbf{r}'
\right]^{1/2}.
	\label{eq:vx_mbj_c_param}
\end{equation} 
In Eq.~(\ref{eq:vx_mbj}), $n$ is the electron density, $\tau$ is the kinetic-energy density and $v^\text{BR}_{\text{x}}$ is the Becke-Roussel potential~\cite{beckePRA1989}.
$\alpha=-0.012$ and $\beta=1.023$ Bohr$^{1/2}$ are parameters which were fitted specifically for band gaps.
%Complications arise from the definition of $c$ when considering the implementation of this functional in pseudopotential codes.
%
Originally, mBJLDA was implemented in an all-electron code, meaning that the quantities in Eqs.~\eqref{eq:vx_mbj} and (\ref{eq:vx_mbj_c_param}) are defined with respect to the total density $n$ of the system, comprising core and valence electrons.
%where the $| \nabla n ({\mathbf{r}}) | / n ({\mathbf{r}}) $ diverges at the positions of the nuclei.
%
However, in the case of a pseudopotential code, such as {\sc ABINIT}, one has to work with a pseudodensity $n$ that differs significantly from the total density in the regions around the atoms (i.e., it is much smoother). As such, it is doubtful that all-electron and pseudopotential calculations of 
Eq.~\eqref{eq:vx_mbj_c_param} should give the same result. In fact, at the very least, the coefficients $\alpha$ and $\beta$ should be reoptimized for pseudodensities (see, e.g., Ref.~\cite{TraorePRB19} for such a procedure with {\sc ABINIT}).

There is also some incongruousness within the {\sc VASP} implementation of mBJLDA. Even if theoretically the PAW method allows to recover the true density of the system, {\sc VASP} performs an additive separation of $c$. This is certainly better than ignoring the contribution of the core density to $c$, but it is a numerical approximation. In practice this leads to an underestimation of $c$ (for most cases, see Fig.~S1) and therefore of the band gap (since the band gaps increases in a monotonous way with $c$). In order to illustrate more concretely the effect of an inaccurate value of $c$ on the band gap, we show in Table~\ref{tab:mbj} the mBJLDA band gap obtained with {\sc VASP} and {\sc WIEN2k} for some of the materials where the discrepancy between the two codes is the largest (from 1 to 4~eV). However, if the {\sc VASP} calculations are done by fixing the value of $c$ to the one obtained from the {\sc WIEN2k} calculation, then the band gap is clearly much closer to the {\sc WIEN2k} value. For instance, for Ne the {\sc VASP} error gets reduced from about 4~eV to less than 1~eV. This shows that a large portion of the error with {\sc VASP} is due to an inaccurate calculation of $c$.

The definition of the $c$ parameter also creates some complications for the generation of a mBJLDA pseudopotential.
As seen in Eq.~\eqref{eq:vx_mbj_c_param}, this parameter becomes ill-defined for (semi-)finite systems. This problem can however be bypassed. For example, Bart\'{o}k and Yates~\cite{BartokPRB19} have recently proposed using a constant value for $c$ during the pseudopotential generation, obtaining good agreement with all-electron calculations.
Another possible solution is the use of the localized version of the mBJLDA potential recently proposed in Ref.~\cite{Rauch2020}.

\begin{table}
\caption{\label{tab:mbj}mBJLDA band gaps (in eV) obtained with {\sc VASP} and {\sc WIEN2k}. The second set of {\sc VASP} results were obtained with the parameter $c$ in Eq.~(\ref{eq:vx_mbj}) fixed to the value obtained from the {\sc WIEN2k} calculation.}
\begin{tabular}{lccc} \hline  \hline 
Solid & {\sc VASP} & {\sc VASP}($c=c_{\text{\sc WIEN2k}}$) & {\sc WIEN2k} \\ 
\hline 
\ce{Ne}    & 18.85 & 21.97 & 22.72 \\
\ce{Ar}    & 12.55 & 13.38 & 13.90 \\
\ce{LiF}   & 12.08 & 12.51 & 12.82 \\
\ce{KCl}   &  7.78 &  8.28 &  8.56 \\
\ce{Al2O3} &  7.86 &  7.96 &  8.31 \\
 \hline  \hline 
\end{tabular}
\end{table}

\section{Conclusions}
\label{sec:conclusions}

In summary, we performed a series of band gap calculations on a test set of 473 materials using all-electron, norm-conserving pseudopotential, and PAW methods. Our goal was to estimate the error on the band gap when standard LDA/PBE pseudopotentials or PAW setups are used inconsistently to perform electronic structure calculations using other density functionals. From our results we concluded the following.
\begin{enumerate}
    \item As expected, consistent pseudopotential calculations are perfectly suited for the evaluation of band gaps. The errors with respect to the all-electron reference results ($\text{MAE}\sim0.02$~eV) are considerably smaller, by one or two orders of magnitude, than the errors (with respect to experiment) due to the approximation to the xc functional.
    \item Using an inconsistent pseudopotential, i.e., one that was generated and optimized for another xc functional,
    increases the mean absolute errors by a factor of three or more, so that the MAE, for instance, can reach values around 0.1~eV. However, this is still smaller than the error due to other theoretical approximations, such as the choice of the xc functional, therefore the pseudopotential approach is still justified for several applications, especially when one is interested in average quantities.
    \item Nevertheless, the error in few specific (and unpredictable) cases can be quite large, with band gaps sometimes differing from the all-electron results by several eV. Therefore, when one requires precise numerical estimations of band gaps, the use of consistent pseudopotentials or all-electron calculations is strongly recommended.
\end{enumerate}

Of course, these conclusions can be relevant not only for semi-local functionals: the use of inconsistent pseudopotentials with hybrid functionals, LDA+$U$, or many-body approaches like $GW$ should also be investigated more thoroughly.

The test set employed here did not contain magnetic materials. Effective-core methods are known to perform sometimes badly for this type of systems (see for example Ref.~\cite{kressePRB2005}). Preliminary calculations indicate that using the wrong pseudopotential in this situation can lead to errors as bad or worse than the ones observed here, but future tests are necessary to be able to properly quantify this effect.

As a final word, band gaps are only one of many quantities of interest for the solid-state community.
Large scale studies of the effect of cross-functional calculations on other properties such as formation energies, geometries, absorption spectra, etc., are important and should be encouraged.
Only with access to such quantitative data we can make informed decisions on the choice of methods for the calculation of electronic properties of materials.

\section{Acknowledgements}
\label{sec:acknowledgements}

J.D. acknowledges support from the Austrian Science Fund (FWF) through project W1243 (Solids4Fun).
M.A.L.M. and S.B. acknowledge partial support from the DFG through the projects TRR 227, SFB 1375, FOR 2857, BO 4280/8-1 and MA 6787/9-1.

\begin{suppinfo}
\begin{itemize}
   \item Spreadsheet containing the dataset and the calculated gaps.
   \item Tables SI-SVII: summary of the relevant statistical quantities for the different error distributions described in the main text.
   \item Figure SI: comparison of $c$ values from mBJ obtained using WIEN2k and VASP.
\end{itemize}
\end{suppinfo}

\bibliography{references}

\end{document}

\title{Supporting Information for 'Validation of pseudopotential calculations for the electronic band gap of solids'}

% List of authors
\author{Pedro M. Borlido}
\affiliation{\jena}
\author{Jan Doumont}
\affiliation{\vienna}
\author{Fabien Tran}
\affiliation{\vienna}
\author{Silvana Botti}
\email{silvana.botti@uni-jena.de}
\affiliation{\jena}
\author{Miguel A.~L. Marques}
\email{miguel.marques@physik.uni-halle.de}
\affiliation{\halle}

\maketitle

\clearpage

\renewcommand{\arraystretch}{0.8}

%%%%%
% Values for LDA@LDA
%%%%%
{
\begin{table}
   \caption{ Statistical data regarding the dispersion of results for LDA calculations with respect to WIEN2k reference values. }
   \label{sup-tab:stats-lda}
   \centering
   \begin{tabular}{l r r r} \hline \hline
                      &        ABINIT:LDA@LDA &        ABINIT:LDA@PBE &       ABINIT:LDA@SLOC \\  \hline
 $\sigma$ (eV)        &                  0.04 &                  0.06 &                  0.13 \\ 
 IQR (eV)             &                  0.03 &                  0.06 &                  0.10 \\ 
 MnE         (eV)     &                  0.00 &                 -0.02 &                  0.01 \\ 
 MADM        (eV)     &                  0.01 &                  0.03 &                  0.05 \\ 
 ME          (eV)     &                  0.01 &                 -0.02 &                 -0.02 \\ 
 MAE         (eV)     &                  0.02 &                  0.05 &                  0.08 \\ 
 MPE (\%) (0.01 eV)   &                  1.96 &                 -1.92 &                  0.08 \\ 
 MPE (\%) (0.25 eV)   &                  0.45 &                 -2.37 &                 -1.26 \\ 
 MPE (\%) (0.50 eV)   &                  0.27 &                 -1.90 &                 -1.13 \\ 
 MPE (\%) (0.75 eV)   &                  0.39 &                 -1.67 &                 -1.00 \\ 
 MAPE (\%) (0.01 eV)  &                  4.91 &                  8.52 &                 10.74 \\ 
 MAPE (\%) (0.25 eV)  &                  2.47 &                  4.92 &                  6.90 \\ 
 MAPE (\%) (0.50 eV)  &                  1.64 &                  3.28 &                  5.11 \\ 
 MAPE (\%) (0.75 eV)  &                  1.40 &                  2.93 &                  4.63 \\ 
 Max. Abs. Err. (eV)  &                 -0.28 &                 -0.29 &                 -1.69 \\ 
             in (\%)  &                -78.24 &                -82.23 &                -26.53 \\ 
 Max. Err. Name       &             \ce{SnTe} &             \ce{SnTe} &             \ce{LaF3} \\ 
 \hline \hline
   \end{tabular}
\end{table}
}
{
\begin{table}
   \addtocounter{table}{-1}
   \caption{ (cont.) Statistical data regarding the dispersion of results for LDA calculations with respect to WIEN2k reference values. }
   \centering
   \begin{tabular}{l r r} \hline \hline
                      &      ABINIT:LDA@HLE16 &          VASP:LDA@LDA \\ \hline 
 $\sigma$ (eV)        &                  0.14 &                  0.04 \\  
 IQR (eV)             &                  0.14 &                  0.04 \\  
 MnE         (eV)     &                 -0.02 &                  0.01 \\  
 MADM        (eV)     &                  0.06 &                  0.02 \\  
 ME          (eV)     &                 -0.06 &                  0.01 \\  
 MAE         (eV)     &                  0.10 &                  0.03 \\  
 MPE (\%) (0.01 eV)   &                 -3.58 &                  2.91 \\  
 MPE (\%) (0.25 eV)   &                 -4.82 &                  0.93 \\  
 MPE (\%) (0.50 eV)   &                 -3.90 &                  0.45 \\  
 MPE (\%) (0.75 eV)   &                 -3.80 &                  0.39 \\  
 MAPE (\%) (0.01 eV)  &                 13.21 &                  6.00 \\  
 MAPE (\%) (0.25 eV)  &                  8.92 &                  2.96 \\  
 MAPE (\%) (0.50 eV)  &                  6.49 &                  2.09 \\  
 MAPE (\%) (0.75 eV)  &                  6.02 &                  1.75 \\  
 Max. Abs. Err. (eV)  &                 -0.93 &                 -0.27 \\  
             in (\%)  &                -14.70 &                -74.83 \\  
 Max. Err. Name       &             \ce{LaF3} &             \ce{SnTe} \\  
 \hline \hline
   \end{tabular}
\end{table}
}

%%%%%
% Values for PBE@PBE
%%%%%
{
\begin{table}
   \caption{ Statistical data regarding the dispersion of results for PBE calculations with respect to WIEN2k reference values. }
   \label{sup-tab:stats-pbe}
   \centering
   \begin{tabular}{l r r r} \hline \hline
                      &        ABINIT:PBE@LDA &        ABINIT:PBE@PBE &       ABINIT:PBE@SLOC \\  \hline
 $\sigma$ (eV)        &                  0.06 &                  0.03 &                  0.13 \\ 
 IQR (eV)             &                  0.06 &                  0.03 &                  0.09 \\ 
 MnE         (eV)     &                  0.03 &                  0.00 &                  0.02 \\ 
 MADM        (eV)     &                  0.03 &                  0.01 &                  0.04 \\ 
 ME          (eV)     &                  0.04 &                  0.01 &                  0.01 \\ 
 MAE         (eV)     &                  0.05 &                  0.02 &                  0.07 \\ 
 MPE (\%) (0.01 eV)   &                  7.50 &                  2.58 &                  4.95 \\ 
 MPE (\%) (0.25 eV)   &                  3.88 &                  0.82 &                  1.68 \\ 
 MPE (\%) (0.50 eV)   &                  2.94 &                  0.44 &                  1.12 \\ 
 MPE (\%) (0.75 eV)   &                  2.49 &                  0.46 &                  0.84 \\ 
 MAPE (\%) (0.01 eV)  &                  8.81 &                  4.72 &                 11.22 \\ 
 MAPE (\%) (0.25 eV)  &                  4.88 &                  2.31 &                  6.50 \\ 
 MAPE (\%) (0.50 eV)  &                  3.80 &                  1.63 &                  5.10 \\ 
 MAPE (\%) (0.75 eV)  &                  3.11 &                  1.35 &                  4.20 \\ 
 Max. Abs. Err. (eV)  &                  0.29 &                  0.19 &                 -1.87 \\ 
             in (\%)  &                 35.52 &                 63.39 &                -29.33 \\ 
 Max. Err. Name       &              \ce{ZnO} &             \ce{GeAs} &             \ce{LaF3} \\ 

 \hline \hline
   \end{tabular}
\end{table}
}
{
\begin{table}
   \addtocounter{table}{-1}
   \caption{(cont.) Statistical data regarding the dispersion of results for PBE calculations with respect to WIEN2k reference values. }
   \centering
   \begin{tabular}{l r r} \hline \hline
                      &      ABINIT:PBE@HLE16 &          VASP:PBE@PBE \\  \hline
 $\sigma$ (eV)        &                  0.11 &                  0.04 \\ 
 IQR (eV)             &                  0.10 &                  0.04 \\ 
 MnE         (eV)     &                  0.00 &                  0.01 \\ 
 MADM        (eV)     &                  0.04 &                  0.02 \\ 
 ME          (eV)     &                 -0.02 &                  0.01 \\ 
 MAE         (eV)     &                  0.07 &                  0.03 \\ 
 MPE (\%) (0.01 eV)   &                 -0.12 &                  3.87 \\ 
 MPE (\%) (0.25 eV)   &                 -1.06 &                  1.17 \\ 
 MPE (\%) (0.50 eV)   &                 -1.53 &                  0.80 \\ 
 MPE (\%) (0.75 eV)   &                 -1.27 &                  0.57 \\ 
 MAPE (\%) (0.01 eV)  &                  9.91 &                  6.01 \\ 
 MAPE (\%) (0.25 eV)  &                  5.87 &                  2.74 \\ 
 MAPE (\%) (0.50 eV)  &                  4.64 &                  1.96 \\ 
 MAPE (\%) (0.75 eV)  &                  3.84 &                  1.59 \\ 
 Max. Abs. Err. (eV)  &                 -1.12 &                 -0.25 \\ 
             in (\%)  &                -17.55 &                 -3.92 \\ 
 Max. Err. Name       &             \ce{LaF3} &             \ce{LaF3} \\ 
 \hline \hline
   \end{tabular}
\end{table}
}

%%%%%
% Values for HLE@PBE
%%%%%
{
\begin{table}
   \caption{ Statistical data regarding the dispersion of results for HLE16 calculations with respect to WIEN2k reference values. }
   \label{sup-tab:stats-hle16}
   \centering
   \begin{tabular}{l r r r} \hline \hline
                      &      ABINIT:HLE16@LDA &      ABINIT:HLE16@PBE &       ABINIT:HLE16@SLOC \\ \hline 
 $\sigma$ (eV)        &                  0.16 &                  0.14 &                  0.13 \\ 
 IQR (eV)             &                  0.16 &                  0.13 &                  0.12 \\ 
 MnE         (eV)     &                  0.04 &                  0.03 &                  0.03 \\ 
 MADM        (eV)     &                  0.07 &                  0.06 &                  0.06 \\ 
 ME          (eV)     &                  0.09 &                  0.06 &                  0.03 \\ 
 MAE         (eV)     &                  0.12 &                  0.10 &                  0.10 \\ 
 MPE (\%) (0.01 eV)   &                  7.39 &                  4.45 &                  4.66 \\ 
 MPE (\%) (0.25 eV)   &                  5.67 &                  3.57 &                  2.62 \\ 
 MPE (\%) (0.50 eV)   &                  5.04 &                  3.28 &                  2.01 \\ 
 MPE (\%) (0.75 eV)   &                  4.26 &                  2.80 &                  1.63 \\ 
 MAPE (\%) (0.01 eV)  &                  9.46 &                  7.42 &                  8.27 \\ 
 MAPE (\%) (0.25 eV)  &                  7.79 &                  6.50 &                  6.28 \\ 
 MAPE (\%) (0.50 eV)  &                  6.72 &                  5.43 &                  5.43 \\ 
 MAPE (\%) (0.75 eV)  &                  5.73 &                  4.62 &                  4.69 \\ 
 Max. Abs. Err. (eV)  &                  0.87 &                  1.10 &                 -0.67 \\ 
             in (\%)  &                 22.92 &                 28.92 &                -17.70 \\ 
 Max. Err. Name       &             \ce{LaF3} &             \ce{LaF3} &             \ce{LaF3} \\ 
   \hline \hline
   \end{tabular}
\end{table}
}
{
\begin{table}
   \addtocounter{table}{-1}
   \caption{(cont.) Statistical data regarding the dispersion of results for HLE16 calculations with respect to WIEN2k reference values. }
   \centering
   \begin{tabular}{l r r} \hline \hline
                      &    ABINIT:HLE16@HLE16 &        VASP:HLE16@PBE \\  \hline
 $\sigma$ (eV)        &                  0.04 &                  0.17 \\ 
 IQR (eV)             &                  0.04 &                  0.13 \\ 
 MnE         (eV)     &                  0.01 &                 -0.01 \\ 
 MADM        (eV)     &                  0.02 &                  0.07 \\ 
 ME          (eV)     &                  0.01 &                 -0.03 \\ 
 MAE         (eV)     &                  0.03 &                  0.11 \\ 
 MPE (\%) (0.01 eV)   &                  1.08 &                 -2.06 \\ 
 MPE (\%) (0.25 eV)   &                  0.78 &                 -1.75 \\ 
 MPE (\%) (0.50 eV)   &                  0.67 &                 -1.39 \\ 
 MPE (\%) (0.75 eV)   &                  0.62 &                 -1.45 \\ 
 MAPE (\%) (0.01 eV)  &                  2.63 &                  6.67 \\ 
 MAPE (\%) (0.25 eV)  &                  2.22 &                  6.12 \\ 
 MAPE (\%) (0.50 eV)  &                  1.75 &                  5.43 \\ 
 MAPE (\%) (0.75 eV)  &                  1.54 &                  5.14 \\ 
 Max. Abs. Err. (eV)  &                  0.17 &                 -1.17 \\ 
             in (\%)  &                 52.54 &                -15.31 \\ 
 Max. Err. Name       &             \ce{CaB6} &             \ce{NaCl} \\ 
   \hline \hline
   \end{tabular}
\end{table}
}

%%%%%
% Values for SLOC
%%%%%
{
\begin{table}
   \caption{ Statistical data regarding the dispersion of results for SLOC calculations with respect to WIEN2k reference values. }
   \label{sup-tab:stats-sloc}
   \centering
   \begin{tabular}{l r r r} \hline \hline
                      &       ABINIT:SLOC@LDA &       ABINIT:SLOC@PBE &      ABINIT:SLOC@SLOC \\ \hline
 $\sigma$ (eV)        &                  0.14 &                  0.16 &                  0.05 \\ 
 IQR (eV)             &                  0.10 &                  0.11 &                  0.03 \\ 
 MnE         (eV)     &                  0.02 &                  0.01 &                  0.01 \\ 
 MADM        (eV)     &                  0.05 &                  0.05 &                  0.01 \\ 
 ME          (eV)     &                  0.05 &                  0.03 &                  0.01 \\ 
 MAE         (eV)     &                  0.08 &                  0.09 &                  0.03 \\ 
 MPE (\%) (0.01 eV)   &                  5.27 &                  3.67 &                  2.13 \\ 
 MPE (\%) (0.25 eV)   &                  3.13 &                  1.67 &                  0.59 \\ 
 MPE (\%) (0.50 eV)   &                  2.65 &                  1.42 &                  0.56 \\ 
 MPE (\%) (0.75 eV)   &                  2.35 &                  1.49 &                  0.38 \\ 
 MAPE (\%) (0.01 eV)  &                  7.98 &                  8.55 &                  3.90 \\ 
 MAPE (\%) (0.25 eV)  &                  5.43 &                  5.94 &                  1.94 \\ 
 MAPE (\%) (0.50 eV)  &                  4.62 &                  4.98 &                  1.60 \\ 
 MAPE (\%) (0.75 eV)  &                  3.97 &                  4.23 &                  1.24 \\ 
 Max. Abs. Err. (eV)  &                  1.33 &                  1.54 &                  0.46 \\ 
             in (\%)  &                 51.37 &                 59.42 &                  7.82 \\ 
 Max. Err. Name       &             \ce{LaF3} &             \ce{LaF3} &            \ce{BaCl2} \\ 
   \hline \hline
   \end{tabular}
\end{table}
}
{
\begin{table}
   \addtocounter{table}{-1}
   \caption{(cont.) Statistical data regarding the dispersion of results for SLOC calculations with respect to WIEN2k reference values. }
   \centering
   \begin{tabular}{l r r} \hline \hline
 Functional           &     ABINIT:SLOC@HLE16 &         VASP:SLOC@LDA \\ \hline
 $\sigma$ (eV)        &                  0.12 &                  0.14 \\ 
 IQR (eV)             &                  0.11 &                  0.13 \\ 
 MnE         (eV)     &                 -0.02 &                 -0.01 \\ 
 MADM        (eV)     &                  0.05 &                  0.06 \\ 
 ME          (eV)     &                 -0.02 &                 -0.03 \\ 
 MAE         (eV)     &                  0.08 &                  0.09 \\ 
 MPE (\%) (0.01 eV)   &                  0.56 &                 -0.40 \\ 
 MPE (\%) (0.25 eV)   &                 -1.14 &                 -1.48 \\ 
 MPE (\%) (0.50 eV)   &                 -1.14 &                 -1.64 \\ 
 MPE (\%) (0.75 eV)   &                 -0.80 &                 -1.72 \\ 
 MAPE (\%) (0.01 eV)  &                  7.41 &                  7.83 \\ 
 MAPE (\%) (0.25 eV)  &                  5.37 &                  5.76 \\ 
 MAPE (\%) (0.50 eV)  &                  4.92 &                  5.25 \\ 
 MAPE (\%) (0.75 eV)  &                  4.27 &                  4.87 \\ 
 Max. Abs. Err. (eV)  &                 -0.54 &                 -0.76 \\ 
             in (\%)  &                -29.85 &                -16.13 \\ 
 Max. Err. Name       &             \ce{GeO2} &             \ce{BiF3} \\ 
   \hline \hline
   \end{tabular}
\end{table}
}

%%%%%
% Values for EV93
%%%%%
{
\begin{table}
   \caption{ Statistical data regarding the dispersion of results for EV93 calculations with respect to WIEN2k reference values. }
   \label{sup-tab:stats-ev93}
   \centering
   \begin{tabular}{l r r} \hline \hline
                      &      ABINIT:EV93@PBE &         VASP:EV93@PBE \\  \hline
 $\sigma$ (eV)        &                 0.08 &                  0.04 \\ 
 IQR (eV)             &                 0.08 &                  0.04 \\ 
 MnE         (eV)     &                 0.02 &                 -0.01 \\ 
 MADM        (eV)     &                 0.04 &                  0.02 \\ 
 ME          (eV)     &                 0.03 &                 -0.00 \\ 
 MAE         (eV)     &                 0.06 &                  0.03 \\ 
 MPE (\%) (0.01 eV)   &                 3.41 &                 -0.18 \\ 
 MPE (\%) (0.25 eV)   &                 2.26 &                 -0.18 \\ 
 MPE (\%) (0.50 eV)   &                 1.92 &                 -0.04 \\ 
 MPE (\%) (0.75 eV)   &                 1.60 &                 -0.13 \\ 
 MAPE (\%) (0.01 eV)  &                 6.64 &                  3.82 \\ 
 MAPE (\%) (0.25 eV)  &                 4.51 &                  2.24 \\ 
 MAPE (\%) (0.50 eV)  &                 3.61 &                  1.70 \\ 
 MAPE (\%) (0.75 eV)  &                 3.16 &                  1.43 \\ 
 Max. Abs. Err. (eV)  &                 0.36 &                 -0.28 \\ 
             in (\%)  &                28.35 &                -76.08 \\ 
 Max. Err. Name       &             \ce{ZnO} &             \ce{SnTe} \\ 
\hline \hline
   \end{tabular}
\end{table}
}

%%%%%
% Values for RPBE
%%%%%
{
\begin{table}
   \caption{ Statistical data regarding the dispersion of results for RPBE calculations with respect to WIEN2k reference values. }
   \label{sup-tab:stats-rpbe}
   \centering
   \begin{tabular}{l r r} \hline \hline
                      &       ABINIT:RPBE@PBE &          VASP:RPB@PBE \\ \hline 
 $\sigma$ (eV)        &                  0.04 &                  0.04 \\ 
 IQR (eV)             &                  0.04 &                  0.03 \\ 
 MnE         (eV)     &                  0.01 &                  0.00 \\ 
 MADM        (eV)     &                  0.02 &                  0.02 \\ 
 ME          (eV)     &                  0.01 &                 -0.00 \\ 
 MAE         (eV)     &                  0.03 &                  0.02 \\ 
 MPE (\%) (0.01 eV)   &                  2.81 &                  1.30 \\ 
 MPE (\%) (0.25 eV)   &                  0.98 &                  0.16 \\ 
 MPE (\%) (0.50 eV)   &                  0.71 &                  0.06 \\ 
 MPE (\%) (0.75 eV)   &                  0.66 &                 -0.03 \\ 
 MAPE (\%) (0.01 eV)  &                  6.52 &                  5.20 \\ 
 MAPE (\%) (0.25 eV)  &                  2.89 &                  2.33 \\ 
 MAPE (\%) (0.50 eV)  &                  2.00 &                  1.56 \\ 
 MAPE (\%) (0.75 eV)  &                  1.70 &                  1.33 \\ 
 Max. Abs. Err. (eV)  &                 -0.30 &                 -0.32 \\ 
             in (\%)  &                -84.80 &                -90.31 \\ 
 Max. Err. Name       &             \ce{SnTe} &             \ce{SnTe} \\ 
\hline \hline
   \end{tabular}
\end{table}
}

%%%%%
% Values for mBJLDA
%%%%%
{
\begin{table}
   \caption{ Statistical data regarding the dispersion of results for mBJLDA calculations with respect to WIEN2k reference values. }
   \label{sup-tab:stats-mbj}
   \centering
   \begin{tabular}{l r } \hline \hline
                      &    VASP:mBJLDA@LDA \\  \hline
 $\sigma$ (eV)        &            0.26 \\ 
 IQR (eV)             &            0.10 \\ 
 MnE         (eV)     &           -0.03 \\ 
 MADM        (eV)     &            0.05 \\ 
 ME          (eV)     &           -0.08 \\ 
 MAE         (eV)     &            0.11 \\ 
 MPE (\%) (0.01 eV)   &           -1.75 \\ 
 MPE (\%) (0.25 eV)   &           -2.06 \\ 
 MPE (\%) (0.50 eV)   &           -1.97 \\ 
 MPE (\%) (0.75 eV)   &           -2.13 \\ 
 MAPE (\%) (0.01 eV)  &            5.75 \\ 
 MAPE (\%) (0.25 eV)  &            4.54 \\ 
 MAPE (\%) (0.50 eV)  &            3.98 \\ 
 MAPE (\%) (0.75 eV)  &            3.74 \\ 
 Max. Abs. Err. (eV)  &           -3.87 \\ 
             in (\%)  &          -17.03 \\ 
 Max. Err. Name       &         \ce{Ne} \\ 
\hline \hline
   \end{tabular}
\end{table}
}

\clearpage
%%%%%
% Comparison of c from mBJLDA
%%%%%
\begin{figure}
   \centering
   \includegraphics{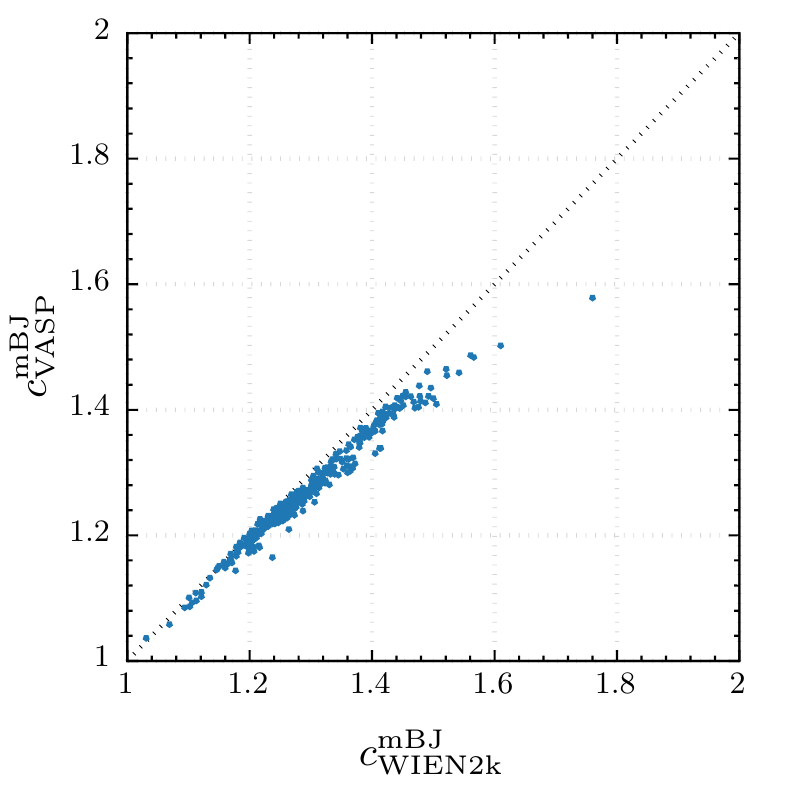}
   \caption{Comparison of computed values of $c$ (see Eq.\eqref{eq:vx_mbj}) for VASP and WIEN2k. }
   \label{sup-fig:comparison-c-mbj}
\end{figure}